# Orbital properties and implications for the initiation of plate tectonics and planetary habitability

*Rajagopal Anand*

Department of Applied Geology, Indian Institute of Technology (Indian School of Mines),
Dhanbad 826004, India

Email: *anandr@iitism.ac.in*

**Abstract**

The existence of plate tectonics on the Earth is directly dependent on the internal viscosity contrast, mass of the planet, availability of liquid water and an internal heat source. However, the initial conditions of rotational velocity and revolutionary periodicity of the Earth around the Sun too must have been significant for the inception of plate tectonics. The initial orbital conditions of the Earth were significantly influenced by the diametrical processes of core segregation and Moon formation and that had probably led to the eventuality of initiation and persistence of plate tectonics. The change in the orbital conditions could have rendered the Earth to evolve in a near-linear trend so that the rotational periodicity of the planet ($T_P$) could approach the time taken for the planet to travel one degree in its orbit around the Sun ($T_{1°}$), that is $T_P \approx T_{1°}$. Such an optimal condition for the rotational and revolutionary periodicities could be essential for the development of plate tectonics on the Earth. This hypothesis has direct implications on the possibility of plate tectonics and life in extrasolar planets and potentially habitable solar planetary bodies such as Europa and Mars.

## 1. Introduction

The initiation and operation of plate tectonics on the Earth is generally attributed to viscosity contrast between the layers of the planet, presence of a continuous heat source and the resultant mantle convection, mass of the planet and availability of liquid water [1–6]. The fundamental factors controlling motion, required for the initiation and persistence of plate tectonics, however have to depend on the rotational as well as the revolutionary movements of the planet. The Earth should have had unique orbital conditions that were modified significantly, few million years after the formation of the Solar System, by the core segregation and Moon formation events. The core segregation process is known to have taken place about 60 Ma after the formation of the Solar System [7–12]. The core segregation process should have increased the rotational velocity of the Earth from its initial spin by the conservation of angular momentum. The formation of the Moon and its separation from the Earth, which is still continuing, has been slowing down the Earth's spin due to tidal drag [13, 14]. This Moon formation event is considered to be either later to core segregation [15, 16], or before core segregation based on the W isotope and Hf/W ratios of the Moon and the Earth [17], or was the trigger and therefore broadly contemporaneous with the core segregation event [7]. In either case, these two events have far reaching implications for the initiation of plate tectonics and the emergence of life. Plate tectonics had provided the essential ingredients in the form of an evolving atmosphere and hydrosphere and changing geomorphology. The latter in turn provided sources and sinks for sedimentation processes that could circulate nutrients for organic evolution and contribute to the ultimate emergence and evolution of life forms as diverse as it is on the Earth.

## 2. Plate Tectonics and Habitability

Looking at the atmospheric composition of the terrestrial planets [18] (Figure 1), it is apparent that unique processes were responsible for the evolution of atmosphere on the Earth. It is logical to assume that the planets falling within the bracket of the supposed 'habitable zone' would have similar atmospheric make up. The present day $CO_2$ levels in the atmospheres of Venus and Mars are close to 96%, while the Earth, sandwiched between these two within the habitability bracket, has an evolved atmospheric $CO_2$ level of < 0.1%. This atmospheric evolution, probably by carbonate weathering and precipitation [19], is a direct consequence of existence of plate tectonics, which happens to be unique on the Earth in the Solar System. An evolving atmosphere had a definitive role to play in the emergence and evolution of Earth-like life forms. Whether the singularity in the occurrence of plate tectonics and life on the Earth - until we make an unequivocal discovery elsewhere - is probabilistic or not remains to be understood. The question that is more enigmatic is how and when plate tectonics began.

*Initial Conditions and the Aftermath*

The initial conditions of the planet-star system are of paramount importance, as those conditions laid the probable path for the planet to evolve to what the Earth is today. Based on empirical data obtained from Proterozoic stromatolites and tidal rhythmites, as well as records of diurnal variations available from the growth patterns of bivalves and brachiopods, the length of the day computed has been increasing upward from about 19 hours a day during end Archaean through about 21 hours a day during Cambrian to the present day length of 23.93 hours a day [20–24]. This increase in the length of the day is attributed to the tidal drag due to the Moon [24].

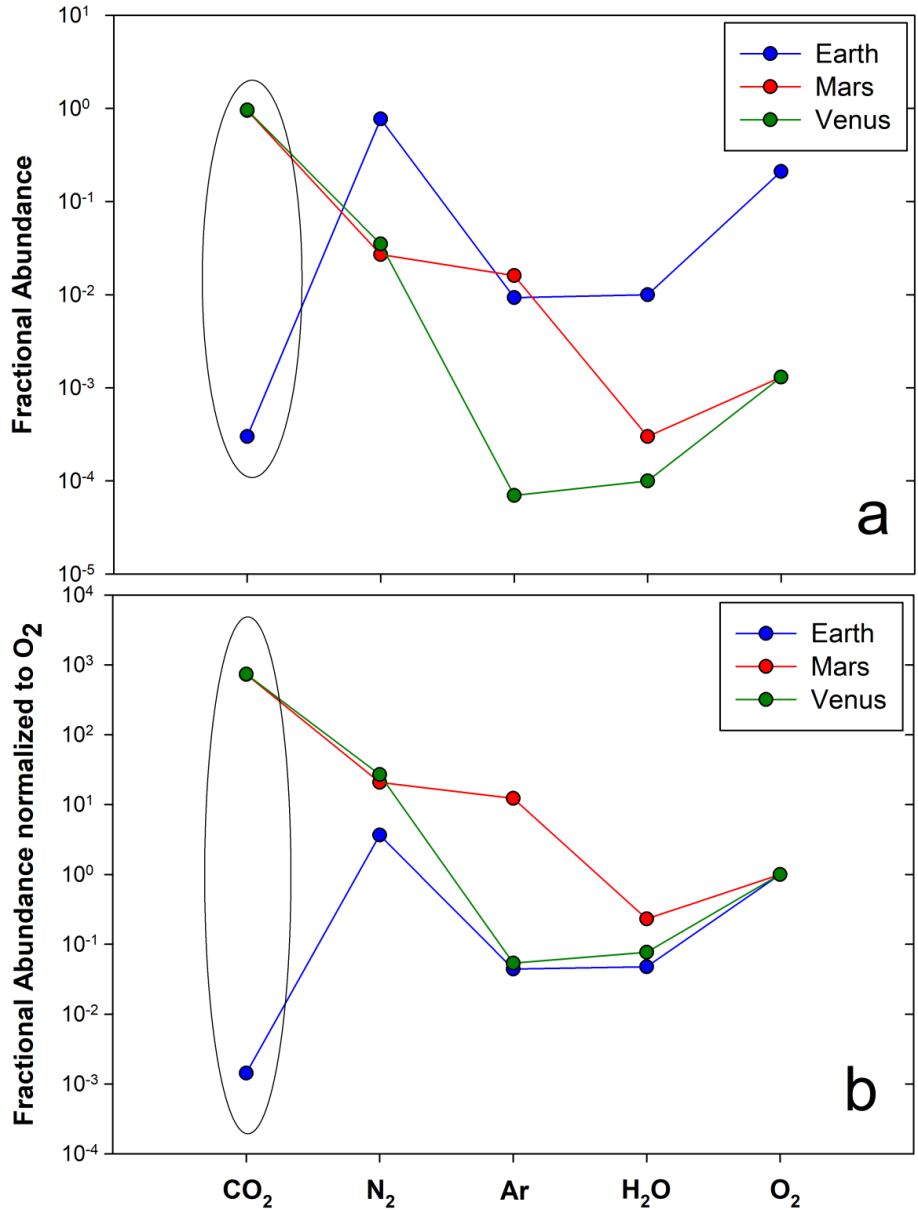

**Figure 1. *a,*** Fractional abundances of atmospheric gases in the terrestrial planets[18]. The $CO_2$ and $N_2$ abundances of the atmospheres of Mars and Venus are nearly identical. These two planets, despite having differing densities and differing distances from the Sun, have retained a similar atmospheric composition for the two most dominant gases. ***b,*** Oxygen normalized fractional abundances of atmospheric gases of terrestrial planets. Earth and Venus have proportionately similar atmospheric makeup for nitrogen, argon, water vapor and oxygen, while the atmospheric levels of $CO_2$, the densest of the above gases, for the Earth have been significantly diminished by processes driven by plate tectonics as compared to Mars and Venus.

The segregation of more than 3,000 km thick core, if it all happened in a single stage during the early phases of the Earth's differentiation, should have increased the rotational velocity of the Earth by the conservation of angular momentum. Varga *et al.* [14] have argued that the length of the day has been non-linear with a major change occurring towards the end of Precambrian. Though supercontinent formation and the distribution of continents and oceans during the end of Precambrian is considered to have caused differential tidal friction, leading to change in the lengths of the day, no significant study supports the assumption. The errors on the paleotide and rhythmite data are not estimated or are expected to be large [24] that a trend for the Proterozoic length of day, different from that for the Phanerozoic, is questionable. As no data on the lengths of days earlier than Proterozoic are available yet, an exact figure on the rotational velocity cannot be arrived at.

### 3. Orbital Properties and Plate Tectonics in Extraterrestrial Bodies

For a planet to develop horizontal tectonics, the primary prerequisite is a continuous source of heat, supplying energy essential for the operation of tectonic movements. This greatly restricts the zone around a star supporting tectonics to the region where the concentration of heavy, as well as, long-lived radioactive elements is greater. The differentiation of the Earth into a Fe-rich core and a silicate mantle provided the right composition for the mantle to further differentiate into an Mg-rich mantle and a silicate crust. Mars and Venus have the right ingredients in the form of heavier elements with long-lived radioactive isotopes and therefore, an additional heat source (apart from the primordial heat) for the operation of mantle convection. Nevertheless, these planets are known to be tectonically dead with a stagnant lid [25, 26] and the occurrence of magmatism is generally attributed to isolated hot spot-like activities [27]. The condition that these planets might not satisfy could be the optimal orbital and spin velocities. Mercury is known to be coupled to the Sun in a 3/2 spin-orbit resonance wherein the planet spins three times about its axis for every two orbits around the Sun [28, 29]. The planet Venus has a sidereal rotation period which is longer than its orbital period by 18 days. These conditions for Mercury and Venus result in a non uniform distribution of the Sun's gravitational force and as a consequence the stress conditions in the interiors of these planets will be non uniform, seriously affecting the possibility for the development of plate tectonics. Habitability in such planets too is doubtful due to the lack of multistability of climate in them [30]. In a similar line of argument it may also be stated that the satellite Europa, which is tidally locked to its planet Jupiter, could not have developed plate tectonics due to uneven stress distribution. In case of the planet Mars, it has an orbital periodicity of 687 days while it has a rotational periodicity of 1.03 days.

## 4. Earth, the Unique Planet

The Earth has a condition where its present day orbital periodicity is 365 days against its rotational periodicity of 0.997 days. It is hypothesized that the orbital conditions of the Earth were set into a dynamic evolution since the completion of the twin events of core segregation and Moon formation. The orbital movements, as observed on the Earth after the twin events, are presumed to be the optimal conditions for the initiation and persistence of plate tectonics. It is predicted that for a planet to develop plate tectonics, these optimal orbital conditions need to be satisfied. The clue could be the near circular movement of the Earth in its orbit, where it moves over one degree in its orbit around the Sun in approximately 24.35 hours and spins about its axis in approximately 23.93 hours (sidereal rotation period) in the present day.

The distance of the Earth from the Sun and its remoteness from a giant planet such as Jupiter permits the Earth to maintain its orbit around the Sun. Also, the eccentricity of the Earth's orbit is low (Table 1) to allow a near circular motion wherein it could attain a constant orbital speed. It is, therefore, safe to assume that the orbital period (and the velocity) of the Earth around the Sun had remained a constant, i.e., 24.35 hours per degree over geologic time. The calculation of one degree considering the orbit either as a circle or as an ellipse does not affect significantly the orbital period due to the very low eccentricity. The rotational periodicities are however likely to have been influenced by the two major processes discussed above as well as by the Late Heavy Bombardment. The change in the initial conditions probably set the evolution of the planet in a near linear trend at some point in time which could eventually reach a condition $T_P \approx T_{1°}$ where, $T_P$ is the sidereal rotational period (SRP) and $T_{1°}$ is the time taken for the planet to travel 1 degree in its orbit. The only planet in the Solar System that satisfies this condition is the Earth (Table 1) and it has taken about 4.6 billion years to achieve this.

It can be argued that any planet might approach this proposed condition over some time. But Mars had acquired its present orbital rotational properties over the same time as the Earth and realizing the proposed condition $T_P \approx T_{1°}$ might not be possible within the lifetime of the Sun. It is also hypothesized that, within a stellar lifetime, when a star has the right essentials to support a planet or planets in a zone, conditioned to hold Earth-like composition and rheology, the $T_P \approx T_{1°}$ condition has to be satisfied within a certain period of the lifetime of the star. In a three dimensional space defined by $T_P$, $T_{1°}$ and age of the planet (Figure 2), the shaded plane represents the constancy in the orbital period of the Earth around the Sun. Data on the probable path of change in the rotational period can be extrapolated on the plane to predict the possibility and time when the planet's rotational period approaches its 1° orbital period. Arriving at a number for this period relative to a star may be implausible. However, observations can be made considering this hypothesis to narrow down possible candidates that could develop plate tectonics and hence, support life.

**Table 1.** Planetary and orbital properties for Solar planets and Europa. Data from http://www.astronomynotes.com

| Planet | Semi-Major Axis (km) | Eccentricity | Orbital Period ($d_E$*) | Orbital Period (hours) | Average Orbital Speed (km/s) | Sidereal Rotation Period (days) | Equatorial Rotation Velocity (km/h) | Distance of 1° (km) | Time taken to travel 1° (hours) | Sidereal Rotation Period (hours) |
|---|---|---|---|---|---|---|---|---|---|---|
| Mercury | 57909100 | 0.2056 | 87.97 | 2111 | 47.87 | 58.65 | 10.89 | 1010704 | 5.86 | 1408 |
| Venus | 108208930 | 0.0068 | 224.7 | 5393 | 35.02 | 243.0 | 6.52 | 1888602 | 14.98 | 5832 |
| Earth | 149598261 | 0.0167 | 365.3 | 8766 | 29.78 | 0.9973 | 1674 | 2610982 | 24.35 | 23.93 |
| Mars | 227939100 | 0.0933 | 687.0 | 16487 | 24.08 | 1.0260 | 868.2 | 3978288 | 45.90 | 24.62 |
| Jupiter | 778547200 | 0.0488 | 4333 | 103982 | 13.07 | 0.4147 | 45300 | 13588212 | 288.8 | 9.95 |
| Saturn | 1433449370 | 0.0565 | 10759 | 258221 | 9.69 | 0.4416 | 35500 | 25018411 | 717.2 | 10.60 |
| Uranus | 2876679082 | 0.0444 | 30685 | 736450 | 6.81 | 0.7183 | 9320 | 50207522 | 2048 | 17.24 |
| Neptune | 4495063661 | 0.0113 | 60190 | 1444560 | 5.43 | 0.6713 | 9660 | 78599920 | 4021 | 16.11 |
| Europa | 670900 | 0.0090 | 3.55 | 85.23 | 13.74 | 3.5512 | 115.1 | 11709 | 0.24 | 85.23 |

* $d_E$ is one Earth day

As an extension, as plate tectonics is very essential for the evolution of a planet's atmosphere, based on the above hypothetical considerations, the probability of life as we know it may be ruled out in potential habitable planetary bodies such as Mars and Europa. Accurate determination of the orbital and rotational periodicities of planets by observation, apart from the other parameters, is therefore essential for predicting the presence of plate tectonics in distant planets.

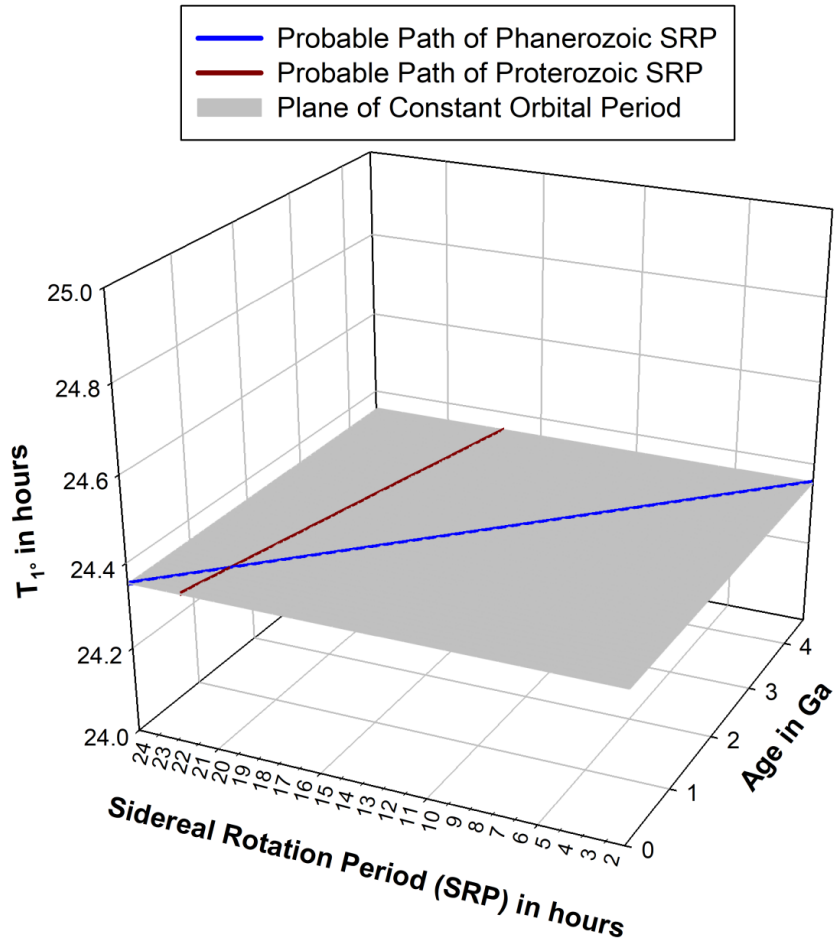

**Figure 2.** The shaded plane represents the constancy of orbital period for the Earth calculated to be 24.35 hours per degree over geologic time. Based on the available data for the lengths of the day [24] the probable paths of the Earth's sidereal rotation period over geologic time extrapolated for Phanerozoic and for Proterozoic are shown on the plane. Every planet with an unperturbed orbit would have a similar plane and depending upon factors such as internal differentiation and external causes such as a retreating satellite, the rotational period of the planet could be influenced.